\documentclass[pre,superscriptaddress,showkeys,showpacs,twocolumn]{revtex4-2}
\usepackage{graphicx}
\usepackage{latexsym}
\usepackage{amsmath}
\begin {document}
\title {Equilibrium behavior in a nonequilibrium system: Ising-doped voter model on complete graph}
\author{Adam Lipowski}
\affiliation{Faculty of Physics, Adam Mickiewicz University, Pozna\'{n}, Poland}
\author{Dorota Lipowska}
\affiliation{Faculty of Modern Languages and Literature, Adam Mickiewicz University, Pozna\'{n}, Poland}
%%%%%%%%%%%%%%%%%%%%%%%%%%%%%%%%%%%%%%%%%%%%%%%%%%%%%%%%%%%%%%%%%%%%%%%%%%%%%
\begin {abstract}
While the Ising model belongs to the realm of equilibrium statistical mechanics, the voter model is an example of a nonequilibrium system.
We examine an opinion formation model, which is a mixture of Ising and voter agents with concentrations $p$ and $1-p$, respectively.  Although in our model for $p<1$ a detailed balance is violated, on a complete graph  the average magnetization in the stationary state for any $p>0$ is shown to satisfy the same equation as for the pure Ising model ($p=1$). Numerical simulations confirm such a behavior, but the equivalence with the pure Ising model apparently holds only for magnetization. Susceptibility in our model diverges at the temperature at which magnetization vanishes, but its values depend on the concentration~$p$. Simulations on a random graph  also show that a small concentration of Ising agents is sufficient to induce a ferromagnetic ordering. 

%For random graphs above the percolation threshold 
\end{abstract}
%\pacs{} 
%	\keywords{migration, Naming Game, language formation}

\maketitle
\section{Introduction}

 Some statistical-mechanics models, initially intended to describe certain physical systems, find numerous applications outside the realm of physics~\cite{loreto}. 
The best example is probably an Ising model, which,  introduced as a model of the magnetic ordering,  recently finds applications in econophysics, computer science, computational biology or neuroscience~\cite{stauffer}. Applications of the Ising model  related to opinion formation are particularly attractive, and a number of versions have been proposed in this context~\cite{watts2007,holyst,sznajd,galam}. 

As is well known~\cite{huang}, the Ising model  undergoes a temperature-driven transition between the ferromagnetic and paramagnetic phases. The transition point is characterized by a singular behavior of some quantities, such as magnetization or susceptibility, but also by vanishing of surface tension. The surface tension in the Ising model remains positive below the transition temperature and is a factor responsible for the curvature-driven coarsening dynamics~\cite{bray}.
Actually, the curvature-driven dynamics is not restricted to  Ising models.  For example, the dynamics of certain opinion-formation models generates an effective surface tension,  as a result  of which such models show certain dynamical similarities with the Ising model~\cite{baron}. There are even some indications that evolution of dialects may  also be subjected to anadopt effective surface tension~\cite{burridge}.

For physicists, a very appealing model of opinion formation is the so-called voter model~\cite{liggett}. The dynamics of the voter model is very simple:  at each step a randomly selected voter adopts an opinion of its randomly selected neighbor. Such tendency to align with the neighbors suggests a similarity to the Ising model, however, some subtle differences result in quite different dynamics of these models. In particular, on two-dimensional lattices, the voter model dynamics is known to be tensionless~\cite{chate} with logarithmically slow coarsening, and on three-dimensional lattices, the voter model does not coarsen at all ~\cite{krapivsky}.  
Let us notice that while the Ising model belongs to equilibrium statistical mechanics, the voter model does not. Indeed, a detailed balance~\cite{grimmet} that characterizes equilibrium systems and that enforces the equal rate of forward and backward processes, does not hold for the voter model. This is because the dynamics of the voter model has configurations, which the system may enter while the exit from which is strictly forbidden \cite{henkel}.

Taking into account the heterogeneity of a human population and the multiplicity of factors affecting opinion-formation processes, a homogeneous model in which each agent acts according to the same rules must certainly be unrealistic. The qualitatively different dynamics of the Ising and the voter models prompted us to examine a model being a mixture of them~\cite{isingdoped}.
In such an Ising-doped voter model, agents are initially classified as Ising or voter type, which determines their dynamical evolution. 
Numerical analysis on two-  and three-dimensional lattices has shown \cite{isingdoped} that even a small fraction of Ising agents is sufficient to induce a ferromagnetic ordering much like in the pure Ising model.  Moreover, the transition temperature in these mixed models decreases upon the decreasing of concentration of Ising agents.

It is easy to realize that in an Ising-doped voter model where both Ising and voter agents are present, certain transitions are also strictly forbidden (while the reverse processes are allowed) and such models are not in equilibrium. Let us also emphasize that violations of detailed balance in nonequilibrium systems result in their more complex behavior such as probability currents, fluxes of particles or energy, or even lack of a stationary steady state.  Their description and understanding, such as that achieved for equilibrium systems, still remains a major challenge~\cite{mukamel,zia,gnesotto}. In the present paper, we examine an Ising-doped voter model on a complete graph. Our study shows that in such a case the magnetization obeys the same equation as in the pure Ising model.  Thus, the system violating a detailed balance  is shown to behave, at least to some extent, as a certain equilibrium system. 

\section{Model}
In our model, at each site~$i$ of a graph, there is an agent, represented as a binary variable $s_i=\pm 1$,  which evolves according to the Ising- or voter-model dynamics. Initially, each agent is assigned the type of dynamics, to which it is subject: with probability~$p$, the agent is set to operate according to the heat-bath Ising dynamics~\cite{newman1999}, and with probability $1-p$, according to the voter dynamics. Our model is thus a quenched mixture of the Ising and voter variables. An elementary step of the dynamics is defined as follows. 
\begin{itemize}
\item  Select an agent, say $i$. 
\item If the variable $s_i$ is of the Ising type, update it according to the heat-bath dynamics, namely, set as +1 with probability 
\begin{equation}
r(s_i\!=\!1)=\frac{1}{1+\exp(-2h_i/T)}, \ \  h_i=\sum_{j_i} s_{j_i},
\label{heat-bath}
\end{equation}
and  as $-1$  with probability $1 - r(s_i\!=\!1)$. \\
The temperature-like parameter~$T$ controls the noise of the system and the summation
in Eq.~(\ref{heat-bath}) is over all neighbors of site~$i$. 
\item If the variable $s_i$ is of the voter type,  select one of its neighbors, say $j$, and set $s_i=s_j$. 
\end{itemize}

We define a unit of time $(t=1)$ as $N$ elementary steps,  where $N$ is the number of sites in the graph.

It is easy to realize that the presence of voter agents ($p<1$) implies a violation of the detailed balance. Indeed, when a voter type agent and all its neighbors are in the same state, then a flip of this agent is strictly forbidden. Since the reversed transition is allowed, it means that the detailed balance in our model (for $p<1$) does not hold.

Let us notice that  models combining two kinds of dynamics have already been examined. For example, Hurtado {\it et al.}~\cite{hurtado}, motivated by the nonequilibrium behavior of interfaces in some disordered systems,  analysed an Ising model with spin variables evolving according to the heat-bath dynamics but with a randomly switched  temperature. In our case, however, the type of the dynamics used by a given agent is initially assigned and kept fixed. 

%%%%%%%%%%%%%%%%%%%%%%%%%%%%%%%%%%%%%%%%%

\subsection{Complete graph}
Our main results are obtained for complete graphs, where each agent is interacting with all the other agents. In the limit $N\rightarrow\infty$, we expect that a mean-field description provides a correct description of the model. Although such a case is already well  known, first let us analyse  the pure Ising model, which corresponds to the case $p=1$.  Moreover, for the purpose of analytical considerations, let us consider a slightly modified dynamics where during a unit of time $t=1$ each site is updated exactly once. 
Assuming that at time~$t$ there are $I_{+}(t)$ and $I_{-}(t)$ Ising agents oriented $+1$ and $-1$, respectively,  we obtain from the dynamical rule~(\ref{heat-bath}) that at time $t+1$ the number of $+1$ agents equals
\begin{equation}
I_{+}(t+1)=\frac{N}{1+\exp[-2(I_{+}(t)-I_{-}(t))/T]}.
\label{evol}
\end{equation} 
In the stationary regime, we might drop the time dependence and introducing $m=(I_{+}-I_{-})/N$ and using $I_{+}+I_{-}=N$, we rewrite Eq.~(\ref{evol}) as
\begin{equation}
\frac{m+1}{2}=\frac{1}{1+\exp(-2mN/T)},
\label{evol_m}
\end{equation} 
or equivalently
\begin{equation}
m=\tanh{(mN/T)}.
\label{evol_tanh}
\end{equation} 
Of course, Eq.~(\ref{evol_tanh}) agrees with the equilibrium statistical-mechanics approach to the Ising model with infinite range of interaction \cite{kadanoff}. 

For $0<p<1$, a fraction $1-p$ of agents operates according to the voter model dynamics. 
As for the evolution of $pN$ Ising-like agents, the analogue of Eq.~(\ref{evol}) now reads
\begin{equation}
I_{+}(t+1)=\frac{pN}{1+\exp[-2(I_{+}(t)+V_{+}(t)-I_{-}(t)-V_{-}(t))/T]},
\label{evol_iv}
\end{equation} 
where $V_{+}(t)$ and $V_{-}(t)$ are the numbers of voter  agents oriented $+1$ and $-1$, respectively. An analogous equation describes the evolution of $I_{-}(t)$ but one can also use the normalization condition $I_{+}(t)+I_{-}(t)=pN$. Evolution equations for $V_{+}(t)$ and $V_{-}(t)$  follow from the rules of the voter model. 
In particular, a probability that a given voter agent at $t+1$ will be set to +1 equals to the fraction of its neighbors that are at time $t$ set to +1, namely $\frac{V_{+}(t)+I_{+}(t)}{N}$. Thus  one can write
\begin{equation}
V_{+}(t+1)=(1-p)N\frac{V_{+}(t)+I_{+}(t)}{N}=(1-p)(V_{+}(t)+I_{+}(t)).
\label{evol_v}
\end{equation} 
 An analogous equation can be written for $V_{-}(t)$, but again the normalization condition $V_{+}(t)+V_{-}(t)=(1-p)N$ can be used. In the stationary regime, Eq.~(\ref{evol_v}) becomes $V_+ =(1-p)(V_++I_+)$  that might be written equivalently as $V_+=\frac{1-p}{p}I_+$ or $V_++I_+=\frac{1}{p}I_+$. Using the  last equation, the stationary form of the evolution equation (\ref{evol_iv}) can be written as
\begin{equation}
 \frac{N}{1+\exp[-2(2I_{+}+2V_{+}-N)/T]}=I_{+}/p= V_++I_+\: .
\label{evol_iv1}
\end{equation} 
With the redefined magnetization 
\begin{equation}
m=\frac{I_{+}+V_+-I_{-}-V_-}{N}=\frac{2(I_{+}+V_+)-N}{N},
\label{newmagnet}
\end{equation}
Eq.~(\ref{evol_iv1}) can be easily shown to be equivalent to Eq.~(\ref{evol_tanh}). Let us notice that for any $0<p\leq 1$, the magnetization $m$ (Eq.~(\ref{newmagnet})) satisfies Eq.~(\ref{evol_tanh}) that is independent of~$p$. Thus, even an arbitrarily small concentration of Ising agents enforces ferromagnetic ordering with the same magnetization as in the pure Ising model (at the same temperature~$T$).

\begin{figure}
\includegraphics[width=\columnwidth]{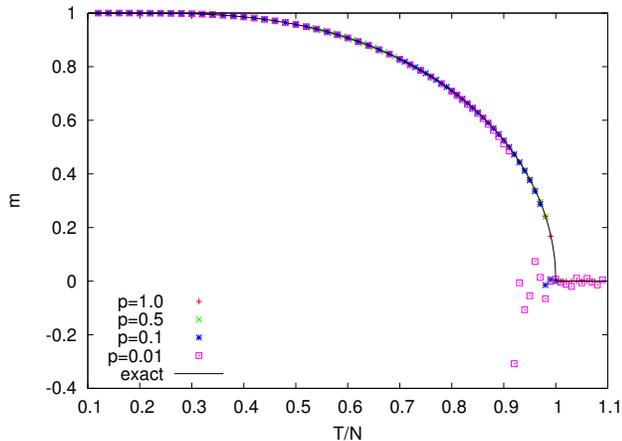}
\vspace{-0mm}
\caption{The temperature dependence of the  magnetization~$m$ for several concentrations of Ising agents $p$. Simulations were made for $N=10^5$ with simulation time  $t=10^6$. 
For each temperature, the initial configuration was ferromagnetic ($s_i=1$).
To reach  the stationary regime, the model relaxed and the relaxation time was equal to the simulation time. The numerical solution of Eq.~(\ref{evol_tanh}) is labelled 'exact'.
}
\label{complete}
\end{figure}

Monte Carlo simulations of our model confirm such a behaviour.  In Fig.~\ref{complete}, we present the temperature dependence of the magnetization~$m$. Let us notice that the critical temperature as calculated from the Eq.~(\ref{evol_tanh}) equals $T=N$. (In some studies \cite{kadanoff}, the Ising model on a complete graph has a Hamiltonian with a coupling constant divided by $N$, which keeps the critical temperature independent of~$N$ but to treat the Ising and voter agents on equal footing, we do not use such a normalization). Simulations for $p=1,\ 0.5$, and $0.1$ (Fig.~\ref{complete}) are nearly not distinguishable and the magnetization indeed seems to vanish at $T/N=1$. Only for $p=0.01$, stronger fluctuations can be seen, especially in the vicinity of the critical point, but we expect that for larger size of the graph~$N$, even in this case the magnetization will vanish at $T/N=1$. Such strong fluctuations are not surprising as  it is the Ising agents that are primarily responsible for the ferromagnetic ordering in our model, and in this case the number of Ising agents is relatively small.
%%%%%%%%%%%%%%%%%%%%%%%%%%%%%%%%%%
%%%%%%%%%%%%%%%%%%%%%%%%%%%%%%%

\begin{figure}
\includegraphics[width=\columnwidth]{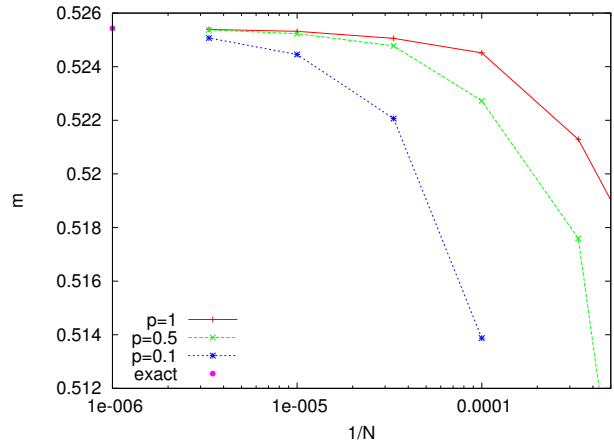}
\vspace{-0mm}
\caption{The size dependence of the magnetization as calculated for $T/N=0.9$ (simulation time  $ t=10^6$).  The solution of Eq.~(\ref{evol_tanh}) for $T/N=0.9$, namely $m=0.52542950\ldots$, is labelled as 'exact'.
}
\label{t09}
\end{figure}

In Fig.~\ref{t09}, it is also demonstrated that the values of magnetization at a given temperature are the same for different~$p$ . Even without any extrapolation,  for $N=3\cdot 10^5$ and $T/N=0.9$, simulations give  $m(p=1)=0.52539(1)$ and $m(p=0.5)=.52537(2)$, which is within  a relative error of $10^{-4}$ with the numerical solution of Eq.~(\ref{evol_tanh}), namely  $m=0.52542950\ldots$ For $p=0.1$  and $N=3\cdot 10^5$, simulations give $m(p=0.1)=0.52507(3)$, which agrees with the expected value within a relative error of $10^{-3}$, but as our data in Fig.~\ref{t09} suggest, with some extrapolation ($N\rightarrow\infty$) a much better agreement could be achieved.  

%%%%%%%%%%%%%%%%%%%%%%%%%%%%%%%%%%
%%%%%%%%%%%%%%%%%%%%%%%%%%%%%%%

\begin{figure}
\includegraphics[width=\columnwidth]{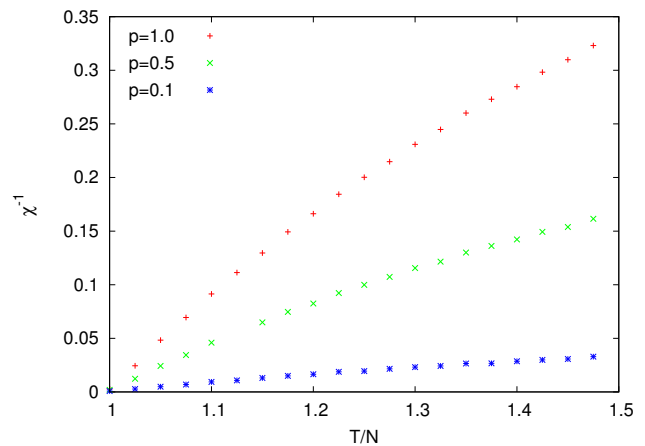}
\vspace{-0mm}
\caption{The temperature dependence of the inverse of susceptibility.  Simulation and relaxation times  were $ t=10^6$  and $N=10^5$.
}
\label{susc}
\end{figure}

We also calculated the variance of magnetization $\chi=\frac{1}{N}(\sum_i s_i)^2$, which for the pure Ising model corresponds, up to the temperature factor, to the susceptibility.  Numerical results indicate (Fig.~\ref{susc}) that $\chi$ diverges at the critical temperature $T/N=1$. Let us notice, however, that the obtained values of $\chi$ depend on the concentration of Ising agents~$p$. It indicates that the equivalence with the $p=1$ case holds but only for magnetization and in general other quantities describing our model most likely depend on~$p$.

%%%%%%%%%%%%%%%%%%%%%%%%%%%%%%%%%%
%%%%%%%%%%%%%%%%%%%%%%%%%%%%%%%

\subsection{Random Graph}
We also examined the behaviour of our model on Erd\"os-R\'enyi random graphs. To generate such a network, for each pair of vertices we place a link with probability $\frac{z}{N-1}$, where $z$ turns out to be the average coordination number of a resulting graph. It is already known that the Ising model placed on such a graph  above the percolation threshold, i.e., for $z>1$, remains ferromagnetic up to the critical temperature \cite{dorog2002,leone2002}
\begin{equation}
T=2/\ln \left(\frac{z+1}{z-1}\right).
\label{rg}
\end{equation}

\begin{figure}
\includegraphics[width=\columnwidth]{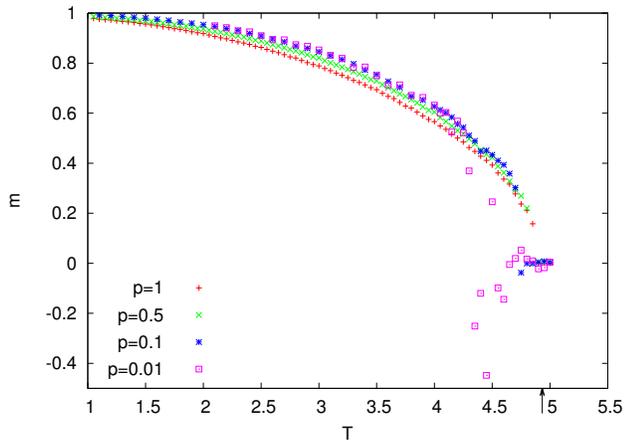}
\vspace{-0mm}
\caption{The temperature dependence of the magnetization for our model on a random graph with the average coordination number $z=5$.  Simulation time  was $ t=10^6$  and $N=10^5$. The arrow indicates the critical temperature for the pure Ising model Eq.~(\ref{rg}).
}
\label{pr5}
\end{figure}

We simulated our model on random graphs with $z=5$. According to Eq.~(\ref{rg}), the pure Ising model ($p=1$) has a phase transition at $T=4.9321\ldots$ Our results show that even for small~$p$, the magnetization deviates only slightly from the case of the pure Ising model $p=1$ (Fig.~\ref{pr5}). Only in the vicinity of the critical temperature, large fluctuations deviate the results from the $p=1$ case, but we expect that for larger~$N$ the good agreement would extend even at larger temperatures. We expect that on finite-$z$ random graphs, the behaviour of magnetization in our model probably depends on~$p$, but numerical results show that this is a rather weak dependance.

%%%%%%%%%%%%%%%%%%%%%%%%%%%%%%%%%%
%%%%%%%%%%%%%%%%%%%%%%%%%%%%%%%

\section{Conclusions}
There is a twofold merit of our work. First, we have shown that a nonequilibrium system with a violated detailed balance can behave, at least with respect to some characteristics,  as  its equilibrium counterpart. Such a result may contribute to a better understanding of the role played by the detailed balance and its violation. 
It would be desirable to examine whether for this kind of models, one can provide an equilibrium statistical-mechanics description, where a certain Hamiltonian would specify a canonical probability distribution of steady state configurations.

In addition, our work shows that even a small fraction of the Ising agents is sufficient to  keep the system in a ferromagnetic state much as in the pure Ising model. On lattices of finite connectivity, numerical simulations show~\cite{isingdoped} a similar effect but with a critical temperature that decreases with the decreasing concentration of the Ising agents. As shown in the present paper, on a complete graph, the critical temperature is independent of this concentration. In the context of opinion formation, we can notice that the Ising dynamics  takes into account a cummulative opinion of the surrounding agents rather than that of a randomly selected neighbor. Our works shows that agents that operate using such a conformity are very influential. An arbitrarily small fraction of them enforces a uniform opinion below the critical temperature but at high temperature they keep the system in the paramagnetic state, i.e., they prevent the voter agents to collapse on one of the absorbing states.

Our simulations suggest that  the Ising-doped voter model on random graphs behaves similarly as on the complete graph, with a magnetization and critical temperature nearly independent of the concentration of the Ising agents. Let us notice that both the Ising~\cite{dorog2002,leone2002} and voter~\cite{castellano2003,sood} models allow for an exact analysis on various heterogenous networks and it is hoped  that a mixture of them could also be similarly studied.

%%%%%%%%%%%%%%%%%%%%%%%%%%

\begin{thebibliography}{}

\bibitem{loreto} C. Castellano, S. Fortunato, and V. Loreto, Statistical physics of social dynamics, Rev. Mod. Phys. {\bf 81}, 591 (2009).

\bibitem{stauffer} D. Stauffer, Social applications of two-dimensional Ising models, Am. J. Phys. {\bf 76}, 470 (2008).

\bibitem{watts2007} D. J. Watts and P. S. Dodds, Influentials, networks, and public opinion formation,  J. Cons. Res. {\bf 34}, 441 (2007).

\bibitem{holyst} J. A. Ho\l yst,  K. Kacperski, and F. Schweitzer, Phase transitions in social impact models of opinion formation, Physica~A {\bf 285}, 199 (2000).

\bibitem{sznajd} K. Sznajd-Weron and J. Sznajd, Opinion evolution in closed community, Int. J. Mod. Phys.~C {\bf 11}, 1157 (2000).


\bibitem{galam} S. Galam, Local dynamics vs.\ social mechanisms: A unifying frame, EPL  {\bf 70} 705 (2005).

\bibitem{huang} K. Huang, {\em Statistical Mechanics}  (New York, Wiley, 1987), 2nd edition.

\bibitem{bray} A. J. Bray, Theory of phase-ordering kinetics, Adv. Phys. {\bf 43}, 357 (1994).

\bibitem{baron} A. Baronchelli, L. DallAsta, A. Barrat, and V. Loreto, 
Topology-induced coarsening in language games, Phys. Rev.~E {\bf 73}, 015102 (2006).

\bibitem{burridge}  Burridge, James, and Tamsin Blaxter, Using spatial patterns of English folk speech to infer the universality class of linguistic copying, Physical Review Research 2.4 : 043053 (2020).

\bibitem{liggett} T. M. Liggett, {\em Interacting Particle Systems} (Springer Verlag, New York, 1985).

\bibitem{chate} I. Dornic, H. Chat\'e, J. Chave, and H. Hinrichsen,   Critical coarsening without surface tension: The universality class of the voter model, Phys. Rev. Lett. {\bf 87}, 045701 (2001).

\bibitem{krapivsky} L. Frachebourg and P. L. Krapivsky, Exact results for kinetics of catalytic reactions, Phys. Rev.~E {\bf 53}, R3009 (1996).

\bibitem{grimmet} Grimmett, G. and Stirzaker, D. (2001). Probability and Random Processes. Oxford University Press.

\bibitem{henkel} J.J. Ramasco  et al., Ageing in the critical contact process: a Monte Carlo study, Journal of Physics A: Mathematical and General {\bf 37}, 10497 (2004).

\bibitem{isingdoped} A. Lipowski, D. Lipowska, and A. L. Ferreira, Phase transition and power-law coarsening in an Ising-doped voter model, Phys. Rev. E {\bf 96}, 032145 (2017).

\bibitem{mukamel} Mukamel D 2000, Phase Transitions in Non-Equilibrium Systems, in Soft and Fragile Matter: Nonequilibrium Dynamics, Metastability and Flow, eds. Cates M E and Evans M R (IOP Publishing, Bristol).

\bibitem{zia} R.K.P Zia and B. Schmittmann, Probability currents as principal characteristics in the statistical mechanics of non-equilibrium steady states, Journal of Statistical Mechanics: Theory and Experiment P07012 (2007).

\bibitem{gnesotto} F.S. Gnesotto et al., Broken detailed balance and non-equilibrium dynamics in living systems: a review,  Reports on Progress in Physics {\bf 81}, 066601 (2018).

\bibitem{newman1999} M. E. J. Newman and G. T. Barkema, Monte Carlo Methods in Statistical Physics Oxford University Press, Oxford, (1999).

\bibitem{hurtado} P. I. Hurtado, P. L. Garrido, and J. Marro, Analysis of the interface in a nonequilibrium two-temperature Ising model, Phys. Rev.~B {\bf 70}, 245409 (2004).

\bibitem{kadanoff} L.P. Kadanoff, Statistical physics: statics, dynamics and renormalization. World Scientific Publishing Company, 2000.

\bibitem{dorog2002} S. N. Dorogovtsev, A. V. Goltsev, and J. F. F. Mendes, Ising model on networks with an arbitrary distribution of connections, Phys. Rev. E {\bf 66}, 016104 (2002).

\bibitem{leone2002} M. Leone, A. Vazquez, A. Vespignani, and R. Zecchina, Ferromagnetic ordering in graphs with arbitrary degree distribution, Eur. Phys. J. B {\bf 28}, 191 (2002).

\bibitem{sood} Sood, Vishal, and Sidney Redner, Voter model on heterogeneous graphs, Physical review letters {\bf 94}, 178701 (2005).

\bibitem{castellano2003} C. Castellano, D. Vilone, and A. Vespignani, Incomplete ordering of the voter model on small-world networks, Europhys. Lett. {\bf 63}, 153 (2003)

\end{thebibliography}
\end {document}